\documentclass{article}

\usepackage{template}

\usepackage[utf8]{inputenc} 
\usepackage[T1]{fontenc}    
\usepackage{hyperref}       
\usepackage{url}            
\usepackage{booktabs}       
\usepackage{amsfonts}       
\usepackage{nicefrac}       
\usepackage{microtype}      
\usepackage{lipsum}
\usepackage{graphicx}
\usepackage{float}
\graphicspath{ {./images/} }

\title{Cluster Analysis of High-Dimensional scRNA Sequencing Data}

\author{
 Jiawei Long \\
  UCLA Fielding School of Public Health\\
  University of California, Los Angeles\\
  Los Angeles, CA 90024 \\
  \texttt{peterljw@g.ucla.edu} \\
   \And
 Yu Xia \\
  UCLA Fielding School of Public Health\\
  University of California, Los Angeles\\
  Los Angeles, CA 90024 \\
  \texttt{xiayu960112@g.ucla.edu } \\
}

\begin{document}
\maketitle
\begin{abstract}
With ongoing developments and innovations in single-cell RNA sequencing methods, advancements in sequencing performance could empower significant discoveries as well as new emerging possibilities to address biological and medical investigations. In the study, we will be using the dataset collected by the authors of \textit{Systematic comparative analysis of single cell RNA-sequencing methods}. The dataset consists of single-cell and/or single nucleus profiling from three types of samples – cell lines, peripheral blood mononuclear cells, and brain tissue, which offers 36 libraries in six separate experiments in a single center. Our quantitative comparison aims to identify unique characteristics associated with different single-cell sequencing methods, especially among low-throughput sequencing methods and high-throughput sequencing methods. Our procedures also incorporate evaluations of every method's capacity for recovering known biological information in the samples through clustering analysis.
\end{abstract}

\section{Introduction}
With the provided dataset of aggregated count matrices of the three samples, we will preprocess each aggregated count matrix into sub count matrices that correspond to separate replicates and sequencing methods. After preprocessing the scRNA sequencing data, we will quantitatively compare and evaluate various methods used on each sample replicate by assessing metrics such as dropout rate and sensitivity measures. Lastly, we will perform various dimensionality techniques along with a number of clustering methods on PBMC and cortex samples to investigate each utilized method’s capacity for recovering known cell types in the samples.

\section{Data Understanding and Preprocessing}
The dataset provides three data files for each sample, which includes the aggregated read count matrix for both replicates and all methods, as well as the gene annotation and cell annotation for the count matrix. The first stage of the study involved dividing the aggregated count matrix into a number of sub count matrices which correspond to different replicates, and sequencing methods and annotating rows with cell names and columns with gene names. 

Among the whole dataset, seven scRNA sequencing methods are present. There are five high-throughput methods, 10x-Chromium-v2, 10x-Chromium-v3, inDrops, Drop-seq, and sci-RNA-seq, alongside two low-throughput methods, Smart-seq2 and CEL-Seq2. By observing the dimensions of the count matrices, low-throughput methods such as Smart-seq2 and CEL-Seq2 tend to be associated with a relatively low number of cells. Figure 1 displays a summary of the count matrices generated after the division of the count matrix for each of the three samples.

\begin{figure} 
    \centering
    \includegraphics[scale=0.83]{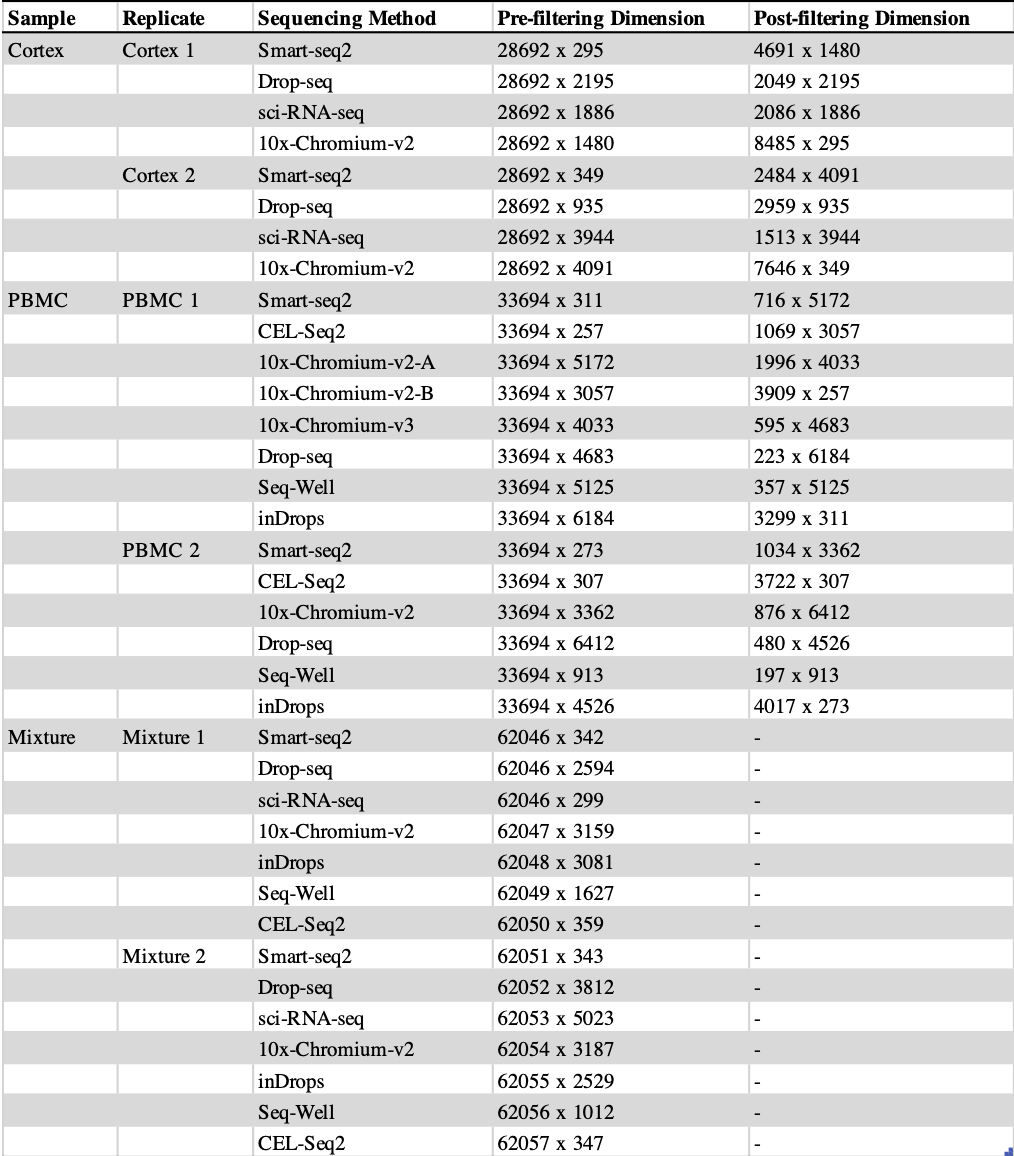}
    \caption{Count Matrices}  
\end{figure}

\section{Comparative Evaluations}

\subsection{Dropout Rate}
A dropout event refers to the occurrence when a transcript is expressed in a cell but is undetected in its mRNA profile. The reason behind the occurrence of dropout events is the low amounts of mRNA in individual cells. The frequency of dropout events depends on scRNA-seq protocols. There exists a general trade-off between the number of cells and the frequency of dropout events, whereas scRNA-seq protocols that generate more cells tend to have a higher frequency of dropout events. Figure 2 demonstrates a graphic illustration of dropout events.

\begin{figure}[H] 
    \centering
    \includegraphics[scale=0.58]{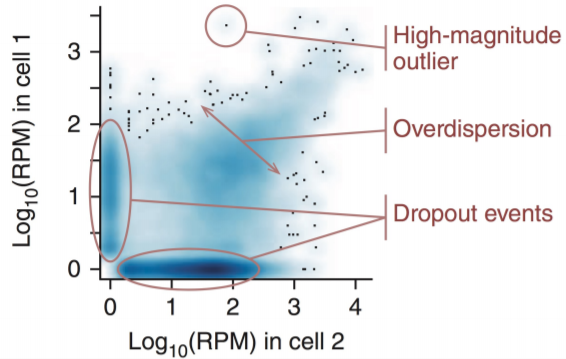}
    \caption{Dimensions of Count Matrices}  
\end{figure}

The dropout rate of a sequencing method indicates the level of the sparsity of a scRNA sequencing method. For each of the count matrix that we have obtained from data preprocessing, we calculated the dropout rate by dividing the number of cells with zero entries by the total number of cells. The outcomes are demonstrated in figure 3 to figure 5 below.

\begin{figure}[H] 
    \centering
    \includegraphics[scale=0.72]{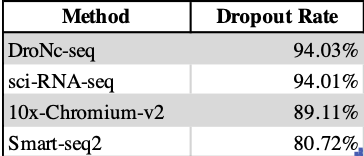}
    \includegraphics[scale=0.72]{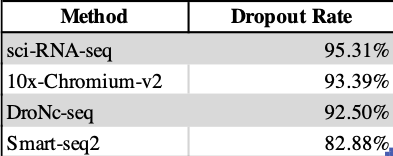}
    \caption{Dropout rates of Cortex 1 (left) and Cortex 2 (right)}  
\end{figure}

\begin{figure}[H] 
    \centering
    \includegraphics[scale=0.72]{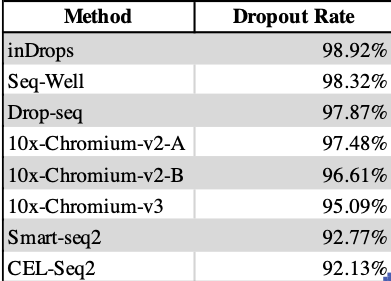}
    \includegraphics[scale=0.72]{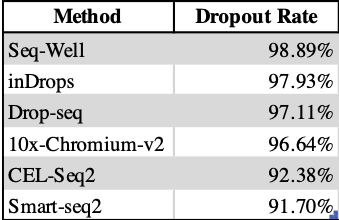}
    \caption{Dropout rates of PBMC 1 (left) and PBMC 2 (right)}  
\end{figure}

\begin{figure}[H] 
    \centering
    \includegraphics[scale=0.72]{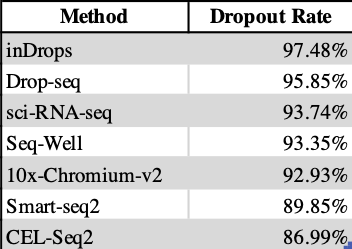}
    \includegraphics[scale=0.72]{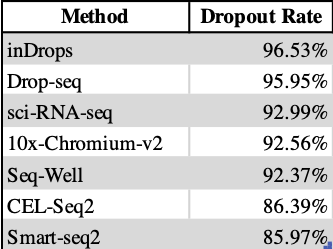}
    \caption{Dropout rates of Mixture 1 (left) and Mixture 2 (right)}  
\end{figure}

According to the dropout rates table,
\begin{itemize}
\item low-throughput plate-based methods such as Smart-seq and Cel-seq are associated with the lowest dropout rates among all methods.
\item high-throughput methods such as Drop-seq, Seq-well, and inDrops have significantly higher dropout rates.
\end{itemize}

\subsection{Sensitivity}
The sensitivity of a scRNA-Seq method refers to the likelihood to capture and convert a particular mRNA transcript present in a single cell into a cDNA molecule present in the library.

To assess the sensitivity of each scRNA sequencing method, we have generated visualizations of two metrics, gene detection per cell and cumulative gene detection. Please refer to figure 6 to figure 9 for the visualisations.
\begin{itemize}
\item Gene detection per cell (left):
The number of genes detected (counts $\geq$ 1) per cell. Each dot represents a cell and each box represents the median and first and third quartiles per replicate and method.
\item Cumulative gene detection (right):
The cumulative number of genes detected as more cells are added. The x-axis represents the number of cells accounted while the y-axis represents the number of genes detected.
\end{itemize}

\begin{figure}[H] 
    \centering
    \includegraphics[scale=0.3]{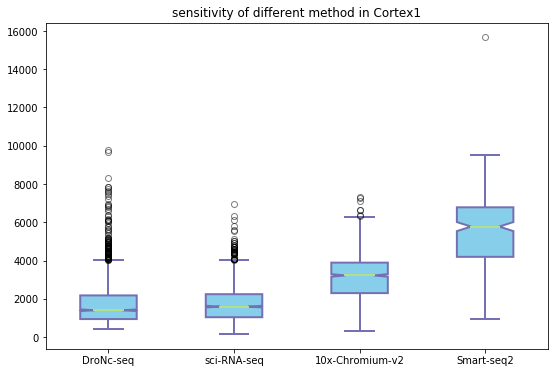}
    \includegraphics[scale=0.4]{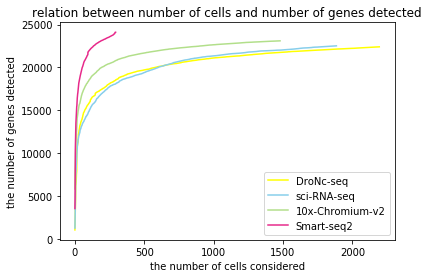}
    \caption{Sensitivity Visualizations for Cortex 1}  
\end{figure}

\begin{figure}[H] 
    \centering
    \includegraphics[scale=0.3]{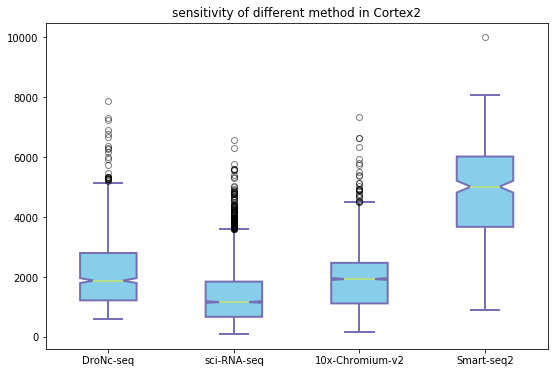}
    \includegraphics[scale=0.4]{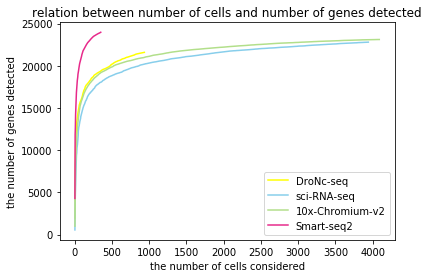}
    \caption{Sensitivity Visualizations for Cortex 2}  
\end{figure}

\begin{figure}[H] 
    \centering
    \includegraphics[scale=0.3]{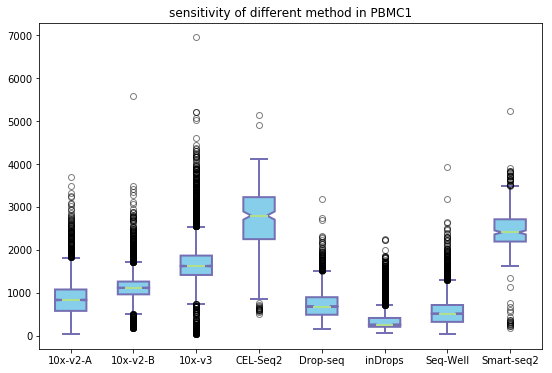}
    \includegraphics[scale=0.4]{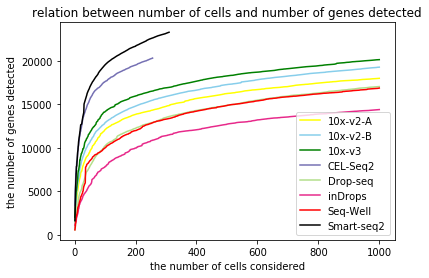}
    \caption{Sensitivity Visualizations for PBMC 1}  
\end{figure}

\begin{figure}[H] 
    \centering
    \includegraphics[scale=0.3]{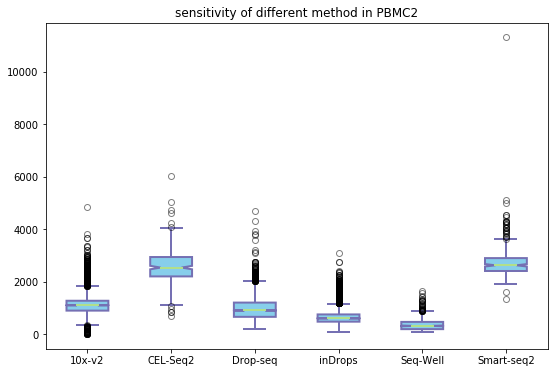}
    \includegraphics[scale=0.4]{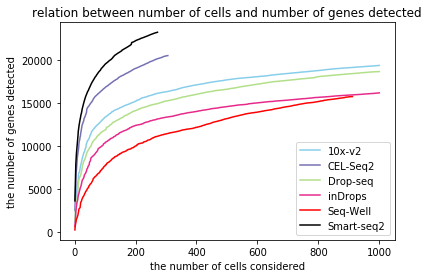}
    \caption{Sensitivity Visualizations for PBMC 2}  
\end{figure}

It is noticeable that low-throughput methods appear to have high sensitivity, in which case they tend to have a higher number of genes detected per cell and they detected the most genes in an efficient manner. In contrast, Drop-seq is consistently associated with low sensitivity across all samples, displaying relatively low gene detection efficiency. Other high-throughput also displayed relatively lackluster performance in terms of sensitivity.

Based on the graphs which display the cumulative number of genes detected with additional cells added, high-throughput methods tend to converge to a relatively low number of detected genes despite increments in the number of cells, implying that there is a group of genes that are potentially difficult to capture with high-throughput methods.

\section{Cluster Analysis}
Sub-population identification, usually via some form of unsupervised clustering, is a fundamental step in the analysis of many single-cell RNA-seq data sets as it helps with identifying underlying gene types. In this section, we have conducted different clustering methods on the  data after applying filtering, normalization, and dimensionality reductions.

\subsection{Filtering}
Although RNA-seq technology has improved the dynamic range of gene expression quantification, low-expression genes may be indistinguishable from sampling noise. The presence of noisy, low-expression genes can lead to poor analysis results. Thus, it is beneficial to remove the low-expression genes in the data before applying clustering techniques. Please refer to figure 1 for dimensions of count matrices after filtering.

We have taken the following two procedures to filter the data.
\begin{itemize}
\item Row sparsity filtering: Remove genes with more than 80\% zero counts.
\item CV filtering: Remove out the bottom 15\% of the remaining genes which have the lowest coefficient of variation. 
\end{itemize}

\subsection{Normalization}
Normalization is an essential step in an RNA-Seq analysis, in which the read count matrix is transformed to allow for meaningful comparisons of counts across samples. In this case, we have applied Quantile normalization. Quantile normalization is a technique for making distributions identical in statistical properties. Each column is converted to rank values before sorting and averaging the ranks to replace the previous ranks.

\subsection{Dimensionality Reduction}
To achieve better and more reasonable clustering visualization results on a two-dimensional space, we needed to apply dimensionality techniques to implement feature engineering.

\paragraph{PCA} Principal component analysis is a linear dimensionality reduction technique, and it performs a linear mapping of the data to a lower-dimensional space in such a way that the variance of the data in the low-dimensional representation is maximized. The major procedure involves calculating the eigenvectors from the covariance matrix, where the eigenvectors that correspond to the largest eigenvalues are used to reconstruct a significant fraction of the variance of the original data. However, in the case of high dimensions, eigenvalues and eigen-vectors of the sample covariance matrix are not consistent, which may affect the performance of PCA. 

\paragraph{t-SNE} t-SNE is a non-linear technique for dimensionality reduction technique. It calculates the probability of similarity of points in high-dimensional space and calculating the probability of similarity of points in the corresponding low-dimensional space, and it minimizes the difference between these conditional probabilities (or similarities) in higher-dimensional and lower-dimensional space for a representation of data points in lower-dimensional space. However, t-SNE is computationally complex and non-deterministic, which makes the technique less robust.

\paragraph{*Projection Pursuit} Projection Pursuit is a component transform technique that looks for a component whose projection vector points to a direction of the quality of interest in data space which can be determined by a Projection Index (PI). In general, projections that go astray more from a normal distribution are viewed as more interesting. As each projection is found, the data are reduced by removing the component along that projection, and the process is repeated to find new projections. The idea of projection pursuit is to locate the projection or projections from high-dimensional space to low-dimensional space that reveal the most insights regarding the structure of the data set. However, one potential issues associated with the technique is that it generally uses random initial conditions to produce projection index components. As a result, when the same projection pursuit is performed on different occasions or by different users, the resulting projection index components are generally not the equivalent. This approach has yet to be implemented in our study.

\subsection{Clustering Methods}

\paragraph{K-means Clustering} K-Means clustering intends to partition n objects into k clusters in which each object belongs to the cluster with the nearest mean. This method produces exactly k different clusters of greatest possible distinction. K-Means clustering minimizes the squared error function with a given k.

\paragraph{Hierarchical Clustering} Hierarchical clustering works by  grouping  the data one by one on the basis of the  nearest distance measure of all the pairwise distance between the data point. Distance could be defined differently and we have used euclidean distance and 1 - correlation to experiment with different distance measurements. Distance calculation methods also vary, which includes single linkage, complete linkage, average linkage, and ward's method (sum of squared euclidean distance is minimized), which is our choice of method in this case. The results from hierarchical clustering will not be displayed due to its relatively poor performance on the data.

From the Single Cell Comparison studies on Single Cell Portal, we acquired the true number of cell types for Cortex and PBMC, respectively.

\begin{itemize}
\item Cortex: 9 cell types
\item PBMC: 9 cell types along with an unidentified cell type
\end{itemize}

Lastly, We used \textbf{silhouette} as a method of interpretation and validation of consistency within clusters of data. The silhouette value is a measure of how similar an object is to its own cluster (cohesion) compared to other clusters (separation). The silhouette ranges from ${-1}$ to ${+1}$, where a high value indicates that the object is well matched to its own cluster and poorly matched to neighboring clusters. Below are the visualizations of the clustering results.

\begin{figure}[H] 
    \centering
    \includegraphics[scale=0.5]{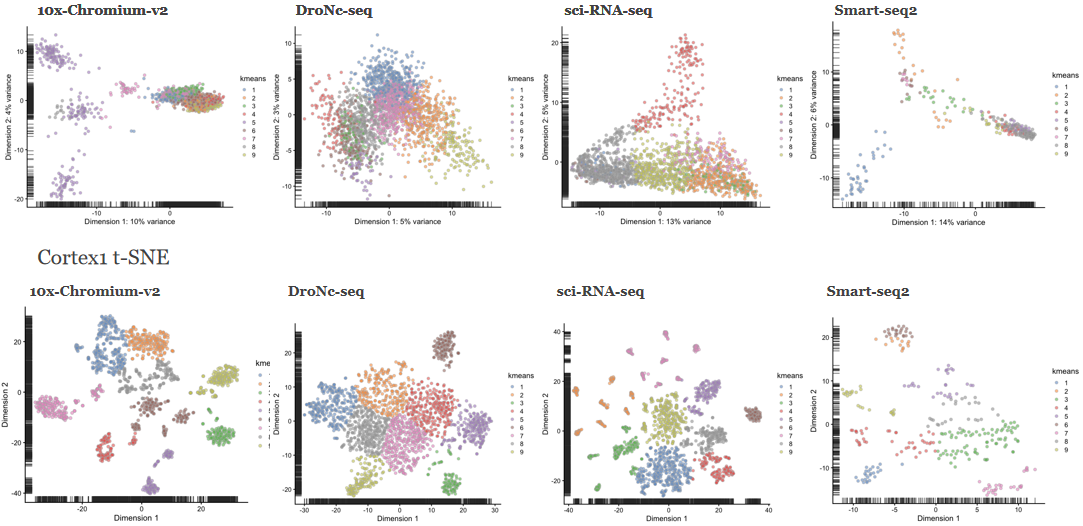}
    \caption{Cortex 1 with PCA and t-SNE}  
\end{figure}

\begin{figure}[H] 
    \centering
    \includegraphics[scale=0.5]{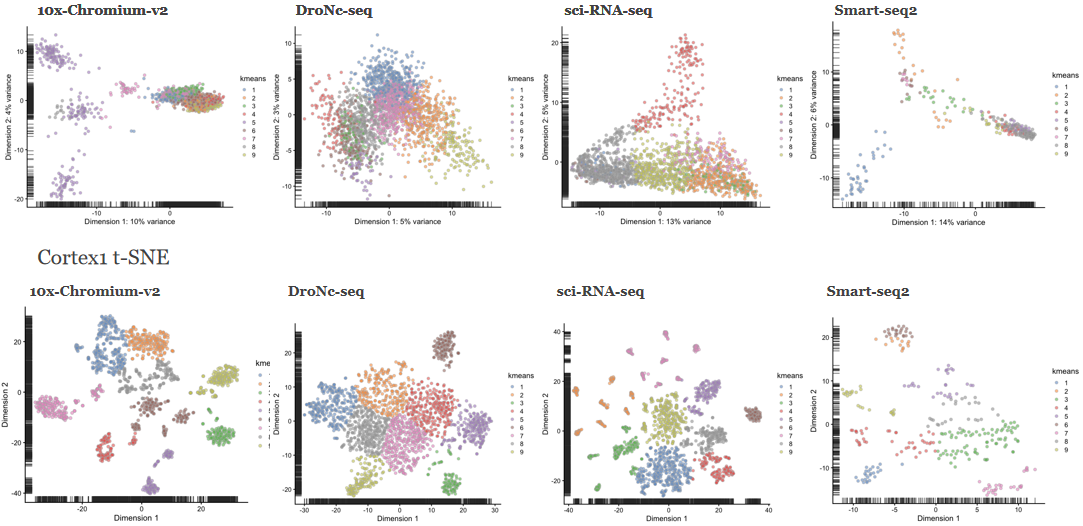}
    \caption{Cortex 2 with PCA and t-SNE}  
\end{figure}

\begin{figure}[H] 
    \centering
    \includegraphics[scale=0.5]{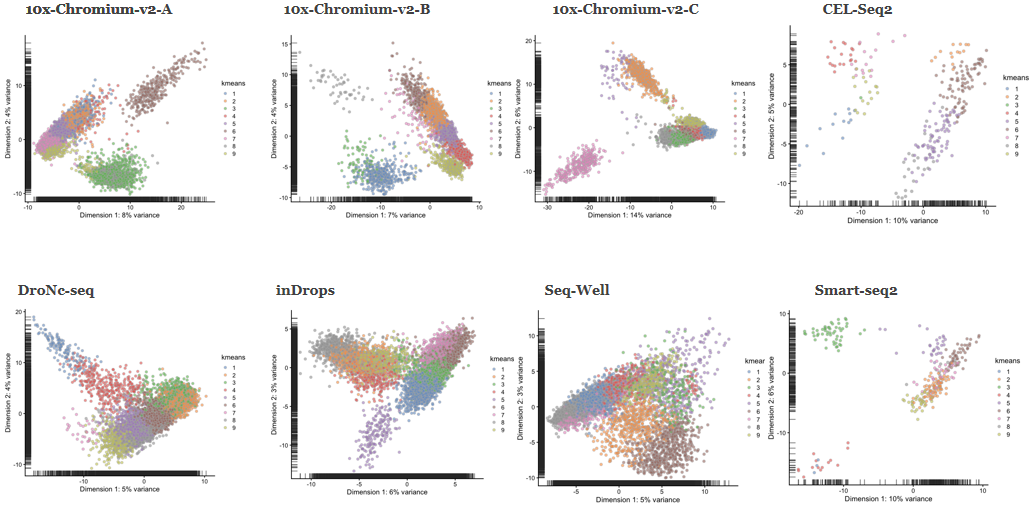}
    \caption{PBMC 1 with PCA}  
\end{figure}

\begin{figure}[H] 
    \centering
    \includegraphics[scale=0.5]{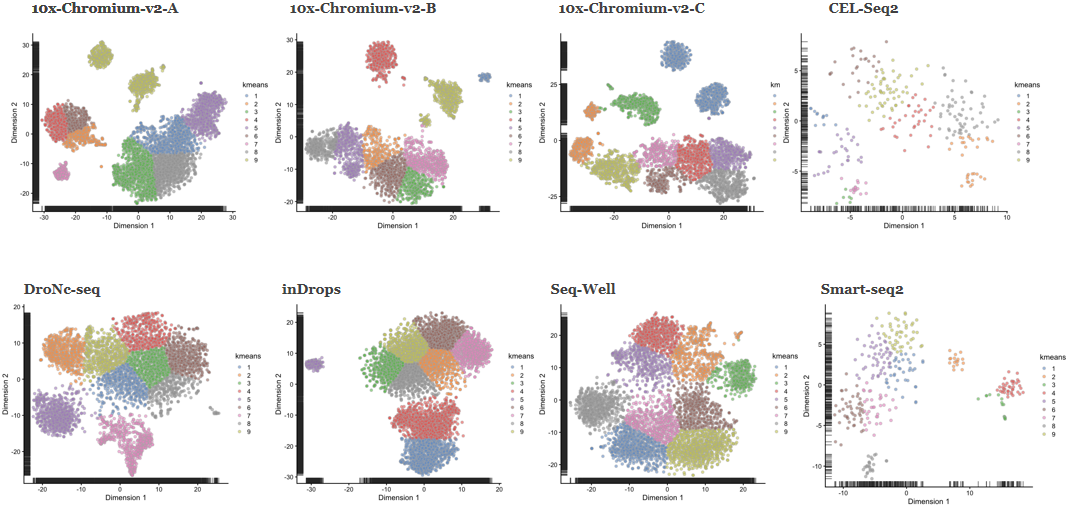}
    \caption{PBMC 1 with t-SNE}  
\end{figure}

\begin{figure}[H] 
    \centering
    \includegraphics[scale=0.5]{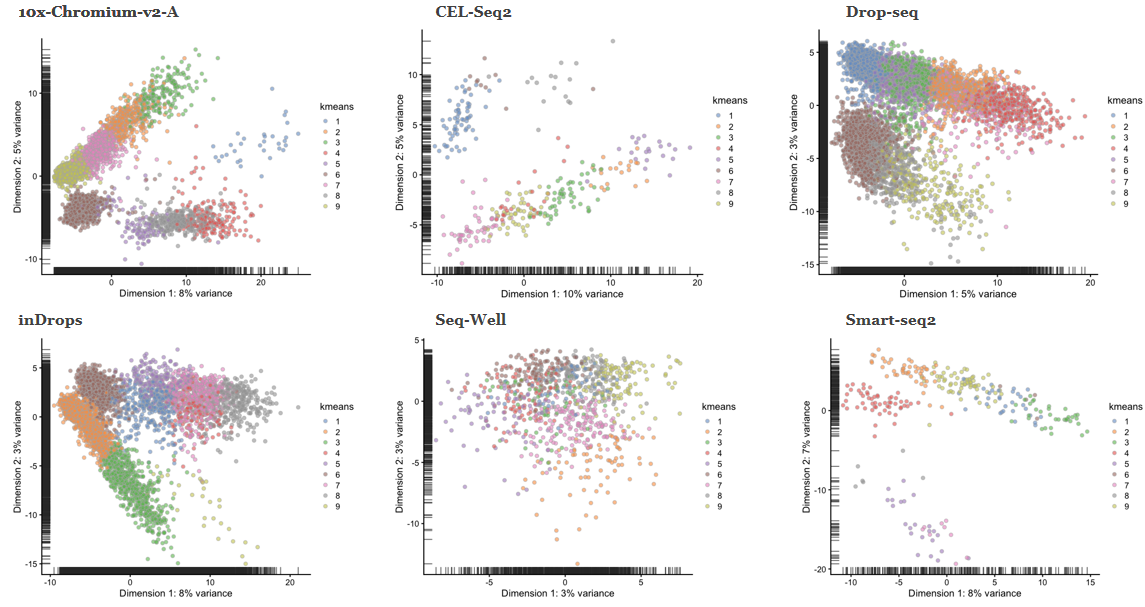}
    \caption{PBMC 2 with PCA}  
\end{figure}

\begin{figure}[H] 
    \centering
    \includegraphics[scale=0.5]{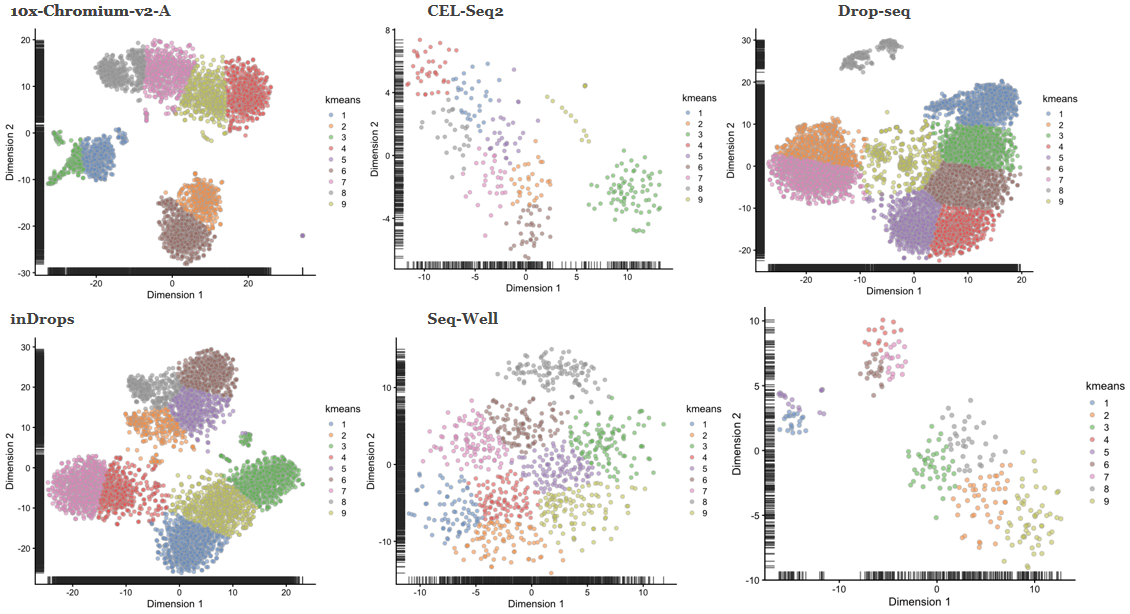}
    \caption{PBMC 2 with t-SNE}  
\end{figure}

\begin{figure}[H] 
    \centering
    \includegraphics[scale=0.8]{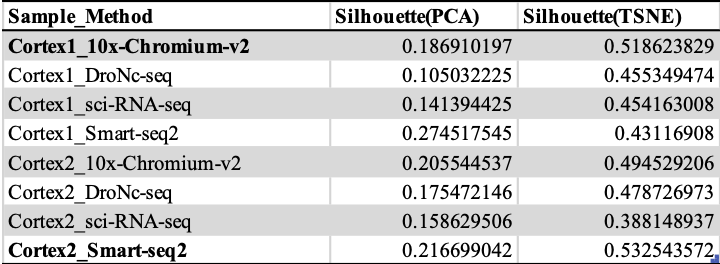}
    \caption{Silhouette Scores for Cortex}  
\end{figure}

\begin{figure}[H] 
    \centering
    \includegraphics[scale=0.8]{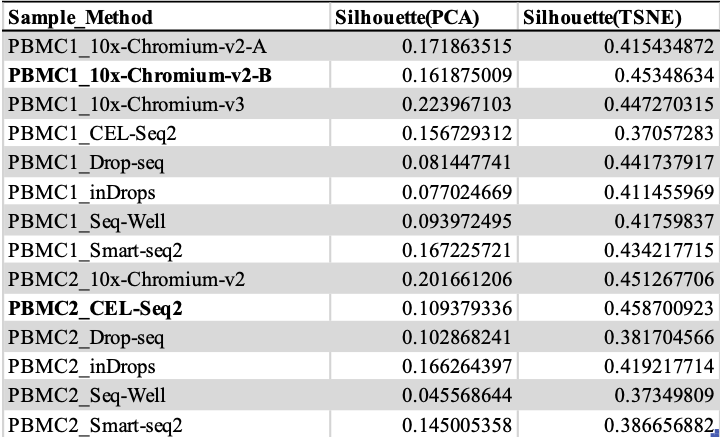}
    \caption{Silhouette Scores for PHBMC}  
\end{figure}

\bibliographystyle{unsrt}

\end{document}